# A CMOS-compatible morphotropic phase boundary

Alireza Kashir*, Hyunsang Hwang*

Center for Single Atom–based Semiconductor Device and Department of Materials Science and Engineering, Pohang University of Science and Technology (POSTECH), Pohang, Republic of Korea

*Email: kashir@postech.ac.kr, hwanghs@postech.ac.kr

**Abstract**

Morphotropic phase boundaries (MPBs) show substantial piezoelectric and dielectric responses, which have practical applications. The predicted existence of MPB in $HfO_2$-$ZrO_2$ solid solution thin film has provided a new way to increase the dielectric properties of a silicon-compatible device. Here, we present a new fabrication design by which the density of MPB $\rho_{MPB}$ and consequently the dielectric constant $\epsilon_r$ of $HfO_2$-$ZrO_2$ thin film was considerably increased. The $\rho_{MPB}$ was controlled by fabrication of a 10-nm [1 nm-$Hf_{0.5}Zr_{0.5}O_2$ (Ferroelectric)/ 1 nm-$ZrO_2$ (Antiferroelectric)] nanolaminate followed by an appropriate annealing process. The coexistence of orthorhombic and tetragonal structures, which are the origins of ferroelectric (FE) and antiferroelectric (AFE) behaviors, respectively, was structurally confirmed, and a double hysteresis loop that originates from AFE ordering, with some remnant polarization that originates from FE ordering, was observed in P-E curve. A remarkable increase in $\epsilon_r$ compared to the conventional $HfO_2$-$ZrO_2$ thin film was achieved by controlling the FE-AFE ratio. The fabrication process was performed at low temperature (250 °C) and the device is compatible with silicon technology, so the new design yields a device that has possible applications in near-future electronics.

**Introduction**

The downscaling of Complementary Metal-Oxide-Semiconductor (CMOS) technology is approaching the limit to reduction in the physical thickness of gate dielectric, without incurring excessive leakage current. However, high-$\kappa$ metal gate transistors require the reduction in equivalent oxide thickness (EOT) of the gate dielectric. Materials with high dielectric permittivity $\epsilon_r$ may enable achievement of this goal. EOT as small as 3.5 - 4 Å has been achieved by using new high $\epsilon_r$ materials [1-4], but they usually have a low band gap $E_g$, and their dielectric properties decrease significantly in thin films, so the task of obtaining satisfactory $\epsilon_r$ in the acceptable thickness range is difficult. Moreover, these materials are mainly incompatible with silicon electronics, or contain toxic elements, and are therefore not appropriate for the current electronics industry.

Many research groups have attempted to reach the ideal region in the graph of $E_g$ versus $\epsilon_r$ [5] by increasing of the dielectric properties of silicon-compatible materials. Methods tried include adding dopants [6-8], stabilizing the metastable high $\epsilon_r$ phase [9], designing nanolaminates [10-11], and preventing formation of an interfacial region [12-13]. Despite the considerable improvements, these attempts seem to have reached their limits, so new approaches must be developed.

Theoretical prediction that morphotropic phase boundaries (MPBs) can form in $HfO_2$-$ZrO_2$ solid solution [14] raises the possibility that ceramic engineering may enable development of lead-free and silicon-compatible high-$\kappa$ dielectric without any degradation in $E_g$, because the chemistries of hafnium and zirconium are more nearly identical than are those of any other two transition metals. Moreover, they have wide band gap (~ 5.5 eV), extreme thinness, good reliability, high binding energy between the oxygen and transition metal ions, and are compatible with CMOS processes. Furthermore, use of $ZrO_2$ and $HfO_2$ in the semiconductor fabrication process is in the mature stage.

An MPB is a region of coexisting phases in the phase diagram of ferroelectrics [15-16]. The MPB separates two competing phases that have distinct symmetries. Realizing MPB usually requires a flat energy surface between different symmetries; this surface facilitates polarization rotation that arises from the coupling between two equivalent energy states, and thus induces large dielectric and piezoelectric responses as a result of enhanced polarizability [17-20]. The MPB has potential applications because the variable that drives the transition (i.e., compositions) is inherent, so the engineering of a ferroelectric (FE) system near MPB may further boost the $\epsilon_r$ without degrading the band-gap.

The existence of MPB has been predicted in $Hf_{1-x}Zr_xO_2$ (HZO) system [14], but the coexistence of tetragonal (t) and orthorhombic (o) phases is substantially affected by fabrication conditions [21-25] (e.g., film thickness, electrode materials, grain size, point defects, annealing conditions) and the energy landscape cannot be simply predicted or controlled. Even though the existence of MPB was predicted at $x = 0.7$ (Ni *et al.* [14]), the increase in $\epsilon_r$ was observed experimentally at $x \approx 0.5$ (Park *et al.* [26]).

Here, we introduce a new design of $HfO_2$-$ZrO_2$ thin film to control and maximize the formation of MPB regions. The $ZrO_2$ thin film shows antiferroelectric (AFE) behavior that originates from a t- (nonpolar) to o- (polar) structural transition induced by the electric field [27]. In contrast, HZO ($x = 0.5$) thin film can exhibit FE properties by stabilization of the metastable non-centrosymmetric o-phase during the thin-film fabrication process [28]. The occurrence of these traits suggests that a nanolaminate of HZO ($x = 0.5$) (FE) and $ZrO_2$ (AFE) followed by an optimum annealing process may lead to formation of MPB at interfacial regions, and yield in a strong dielectric response.

**Experiments**

Three different types of metal-insulator-metal (MIM) capacitors were fabricated on $SiO_2$/Si substrate. They were W/$ZrO_2$/W, W/HZO/W and W/ [HZO (1nm)/$ZrO_2$ (1nm)] × 5 /W (W/HZZ/W). In all three devices the insulator layer was ~ 10 nm thick. The W bottom and top electrodes were deposited using rf-sputtering at room temperature (RT). The HZO and $ZrO_2$ thin films were deposited using atomic layer deposition (ALD). The Hf precursor was tetrakis (ethylmethylamido) hafnium (IV) (Hf [N-($C_2H_5$) $CH_3$]$_4$); the Zr precursor was Tetrakis (ethylmethylamido) zirconium (IV) (Zr [N-($C_2H_5$) $CH_3$]$_4$). The oxidant was $O_3$ (276 g/Nm$^3$). During the growth of HZO, the $HfO_2$ and $ZrO_2$ layers were deposited at a ~ 1:1 cycle ratio to form $Hf_{0.5}Zr_{0.5}O_2$ layers. An ozone pulse duration of 15 s was applied for each deposition; this duration appeared to be an optimum dosage to remove carbon contaminants from the deposited materials at a given condition [29]. The substrate temperature was maintained constant at 250 °C during deposition of all insulating films presented in this study, and the precursors' temperature was kept at 90 °C. The growth rates of HfO$_2$ and ZrO$_2$ were almost identical at ~ 1 Å/cycle. The number of HfO$_2$ and ZrO$_2$ cycles required to deposit (1 nm HZO/1 nm ZrO$_2$) × 5 nanolaminates was determined by considering these growth rates. Top electrodes with sizes from 30 × 30 µm$^2$ to 100 × 100 µm$^2$ were patterned using a lift-off process by a photolithography method. Finally, the capacitors with different electrode areas were annealed under 1000 sccm high-purity N$_2$ flow at 500 °C for 30 sec. Moreover, the W/HZZ/W devices have gone through variety of annealing conditions from 300 °C to 700 °C to achieve the optimum dielectric performance.

The crystal structures of the films were investigated using an X-ray diffractometer in grazing incidence geometry (GIXRD) and high-resolution and scanning transmission electron microscopy (HRTEM and STEM). To determine the elemental composition of the films, X-ray photoelectron spectroscopy (XPS) was performed using the depth-profiling mode by Ar$^+$ ion sputtering on as-grown uncapped samples (Figure S1). The ferroelectric properties of the devices were measured using an LCII ferroelectric precision tester (Radiant Technologies), and the capacitance vs electric field (*C-E)* were evaluated using a Keysight B1500A semiconductor device parameter analyzer. The C–E measurement was executed with an amplitude of 50 mV at a frequency of 10 kHz. Rayleigh dielectric measurements were carried out at a frequency of 10 kHz, wherein the ac-field strength was ramped from 1 mV to 2 V. An Agilent E4980A Precision LCR Meter was used and the electric field was increased incrementally by + 5 mV steps. The dielectric constants were extracted from capacitance values obtained from Rayleigh measurements. The measurements were conducted at RT.

**Results and discussion**

A schematic of the W/HZZ/W device and an atomic model are presented in figure 1a and b, respectively. Recent phase-field modeling (Ni *et al.* [14]) studied the effect of Zr concentration (*x*) on the total energy of HZO system. This modeling revealed the total energy of HZO system versus the sub-lattice polarization ($P_1$, $P_2$). The orthorhombic FE phase is energetically favored if the energy minimum lies on the anti-diagonal line, where the sub-lattice polarizations equal each other ($P_1 = P_2$). In contrast, the tetragonal AFE phase is energetically favored if the energy minimum lies on the diagonal line, where the sub-lattice polarizations are opposite to each other ($P_1 = -P_2$). The simulations predicted that a composition between Hf$_{0.5}$Zr$_{0.5}$O$_2$ and ZrO$_2$ (0.5 < *x* < 1) would have a flat energy surface between different symmetries ($P_1 = P_2$ and $P_1 = -P_2$) (as it is illustrated in the figure 1c). This flat energy surface facilitates the polarization rotation between FE and AFE orderings stimulated by small external electrical or mechanical stimuli and results in a substantial increase in dielectric response [14].

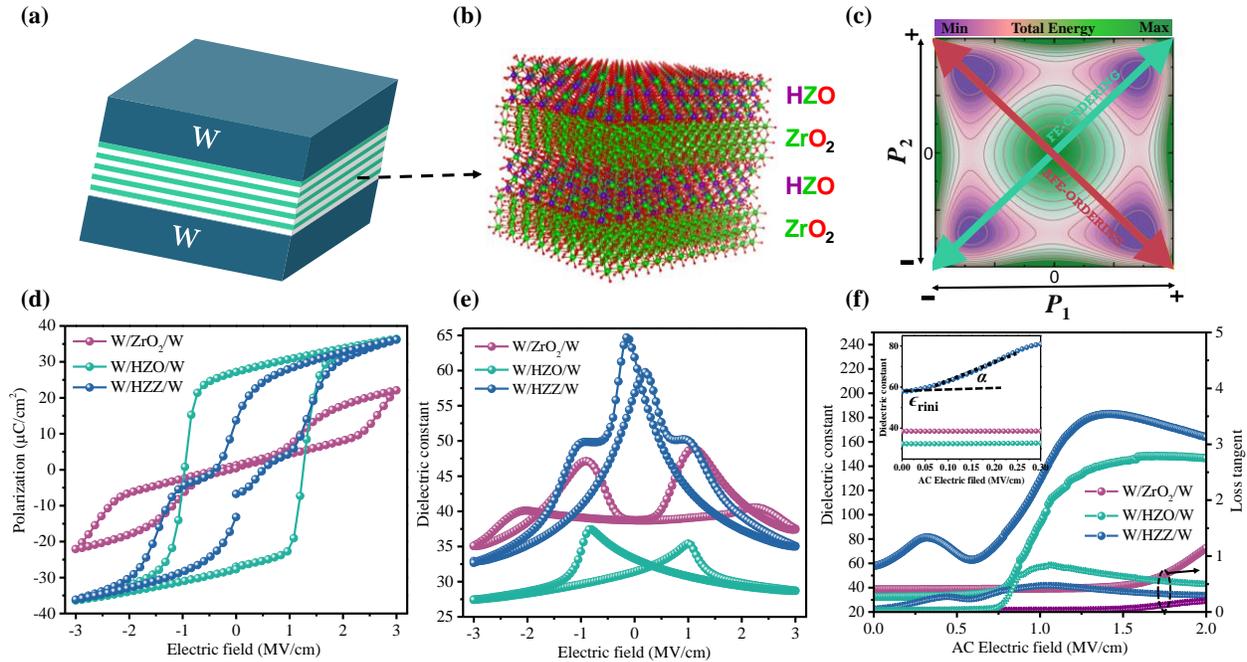

**Figure 1**. (a) Schematic of W/HZZ/W nanolaminate capacitor, (b) an atomic view of HZO/ZrO$_2$ supercell. Purple spheres: Hf, green: Zr, red: O. (c) Total energy contour on the sub-lattice polarization ($P_1$, $P_2$) plane of Zr-doped HfO$_2$ system in a composition between $0.5 < x < 1$ [14]. (d) Polarization vs. electric field $E$, (e) dielectric constants vs. applied biased $E$ and (f) Rayleigh dielectric measurements for W/ZrO$_2$/W, W/HZO/W and W/HZZ/W devices fabricated on SiO$_2$/Silicon substrates and annealed at 500 °C for 30 s.

The curves of polarization versus electric field (*P-E*) of W/ZrO$_2$/W, W/HZO/W and W/HZZ/W devices annealed at 500 °C for 30 s (figure 1d) show that the W/ZrO$_2$/W device has AFE behavior with $2P_r = 0$ and a double-hysteresis loop. The density functional theory calculations predicted a small energy difference of ~ 1 meV/f. u. between the nonpolar t- and polar o-structure in ZrO$_2$ [27]; this difference is characteristic of AFE. In contrast, the HZO sample has FE properties with $2P_r = 54$ μC/cm$^2$ which is believed to originate from the metastable o-phase [28]. The W/HZZ/W device shows a combination of FE and AFE features with $2P_r = 13$ μC/cm$^2$. The maximum polarization $P_{max}$ of HZZ nanolaminate is almost the same as in HZO, and is attributed to the electric field-induced reversible structural change from t- to o-phase. As the electric field was removed, the polarization dropped. A remnant polarization $P_r$ ~ 6.5 µC/cm² in HZZ film originated from the o-phase, which is non-centrosymmetric and thus ferroelectric. Ultrathin Zr-doped HfO$_2$ film with ~ 1-nm thickness maintains its FE properties [30]. Thus, a combination of FE (mainly from HZO layers) and AFE (mainly from ZrO$_2$ layers) emerged in HZZ nanolaminates. This behavior may introduce MPB regions at the FE/AFE interfacial areas. This fact was revealed by the capacitance – electrical field (C-E) measurement, which shows a considerable increase in $\epsilon_r$ of HZZ nanolaminate at bias electric field $E = 0$ MV/cm (figure 1e). The HZO and ZrO$_2$ devices showed $\epsilon_r$-E behavior that agreed well with FE and AFE properties, respectively. At $E = 0$ MV/cm, HZZ had $\epsilon_r = 60$ whereas ZrO$_2$ thin film had $\epsilon_r = 39$, and HZO thin films had $\epsilon_r = 32$. Therefore, the nanolaminate design increased the $\epsilon_r$ of ZrO$_2$-HfO$_2$ thin film by 54%, and achieved the highest $\epsilon_r$ yet reported for an HfO$_2$-ZrO$_2$ device. In ZrO$_2$ devices the maximum $\epsilon_r$ occurred before 0 MV/cm, as is characteristic of AFE behaviors, whereas in HZO devices the maximum $\epsilon_r$ occurred after 0 MV/cm, as is characteristic of FE behaviors. HZZ nanolaminate had maximum $\epsilon_r$ near $E = 0$ MV/cm.

Rayleigh dielectric measurement of HZZ nanolaminate was performed to further investigate its dielectric behavior. The HZO and ZrO$_2$ behaviors were presented as a reference (figure 1f). The Rayleigh measurement studies the behavior of FE materials in three different regimes: low, high, and switching fields [31-33]. In the low-field regime, an enhanced field-independent constant permittivity represented by the reversible Rayleigh parameter $\epsilon_{r_{ini}}$ indicates large intrinsic lattice and reversible motion of the interface around its equilibrium position along the walls of the potential energy well. In the high field (Rayleigh regime), a steep linear slope indicated by the Rayleigh coefficient α is typical of increased irreversible displacement of domain walls or phase boundaries from one potential energy well to another, and acquiring a new position. Thus, the overall dielectric response in the Rayleigh region is represented by [33]

$$\epsilon_r = \epsilon_{r_{ini}} + \alpha E_0 \qquad 1$$

where $E_0$ is the AC electric field amplitude. The Rayleigh parameters ($\epsilon_{r_{ini}}$ and $\alpha$) for nanolaminate device are presented in the inset of figure 1f.

The ZrO$_2$ had a constant $\epsilon_r$ at $E < 1.5$ MV/cm, as expected from the tetragonal non-polar structure (Figure 1f). $\epsilon_r$ increased sharply at $E > 1.5$ MV/cm; this behavior can be attributed to formation of o-phase under a strong electric field. Therefore, a feature similar to the FE system (i.e., HZO) can be observed in ZrO$_2$ film at relatively high $E$. HZO film had $\epsilon_{r_{ini}} \sim 33$, which is smaller than that of ZrO$_2$ ($\epsilon_{r_{ini}} = 39$, (Table 1)). The difference can be understood considering the formation of o-phase in Zr-doped HfO$_2$. $\epsilon_r$ is higher in the t-phase than in the o-phase [34].

The HZO film had a small $\alpha = 118$ µm/kV, because the film has a low domain-wall mobility [31]. At $E > 750$ kV/cm, a sharp increase of $\epsilon_r$ can be observed which corresponds to the onset of switching.

**Table 1.** The values of $\epsilon_{r_{ini}}$ (dielectric constant at low-field regime) and $\alpha$ [µm/kV] for different devices extracted from the Rayleigh dielectric measurements (Figure 1f).

| Parameter | Sample | | |
|---|---|---|---|
| | ZrO$_2$ | HZO | HZZ |
| $\epsilon_{r_{ini}}$ | 39 | 33 | 60 |
| $\alpha$ | 0 | 118 | 847 |

The HZZ nanolaminate had $\epsilon_{r_{ini}} \sim 60$, which is considerably higher than that of HZO and ZrO$_2$. Moreover, in the initial step of the Rayleigh dielectric measurement of HZZ nanolaminate, $\epsilon_r$ increased steeply from 60 to 80 at $E = 300$ kV/cm; this is twice the $\epsilon_r$ of t-ZrO$_2$ thin film.

The HZZ nanolaminate shows two peaks in its dielectric response, whereas HZO and ZrO$_2$ show only one. A non-uniform distribution of the energy barriers causes a smooth blending between the three regions; this observation suggests that reducing the extent of the Rayleigh region, the increase of $\epsilon_r$ at the onset of increased amplitude of AC electric field might point the degenerate energy state of FE and AFE orderings. Compared to the HZO FE film, the Rayleigh region is hardly distinguishable for HZZ film and a relatively steep slope ~ 8 times higher than that of HZO in $\epsilon_r$-$E$ graph was observed.

The cross-sectional STEM study collected on HZZ nanolaminate confirmed that the ALD and post-annealing processes had fabricated the HZO/ZrO$_2$ layer-by-layer structure (Figure 2a).

Indeed, the gray (HZO layer) and dark (ZrO$_2$ layer) regions which present ~ 1-nm HZO-dominant and ~ 1-nm ZrO$_2$-dominant layers, are distinguishable in HAADF-STEM images. It is worth mentioning that the as-deposited amorphous nanolaminate was crystalline through an annealing process at 500 ºC for 30 sec; during crystallization (nucleation and growth), inter-diffusion of adjacent layers (HZO and ZrO$_2$) is inevitable so a sharp interface cannot form. Moreover, we should note that the formation of an interfacial layer between W and HZZ is unavoidable since the HZZ layer was deposited at 250 °C under ozone pulses with a length of 15 sec [29]. Formation of this region is obvious between W and HZZ layers from the STEM image. Especially, the W bottom electrode was substantially affected by fabrication process since the W/HZZ bottom interface was formed during the ALD process.

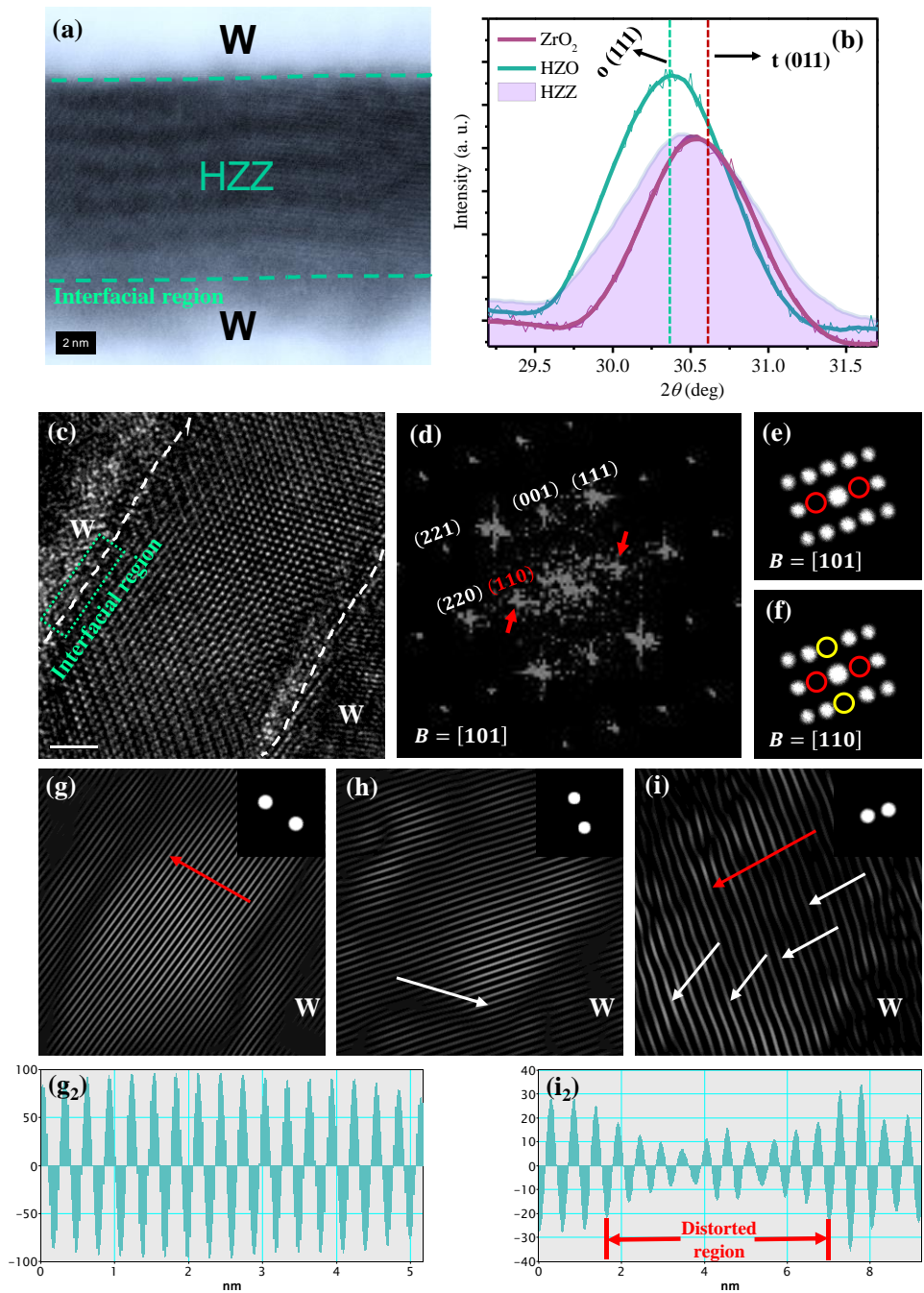

**Figure 2.** (a) HAADF-STEM cross-sectional image from W/HZZ/W device, (b) XRD patterns of $ZrO_2$, HZO and HZZ films. (c) HRTEM cross-sectional image from W/HZZ/W device, (d) the fast Fourier transform (FFT) pattern of the HZZ layer. SAED diffraction simulations of $Pca2_1$ phase along the (e) [101] and (f) [110] zone axes. Along the [110] zone, we note that both the (001) and the (1-10) spots are forbidden. Along the [101] zone, (10-1) is forbidden. (g) The filtered (111), (h) (001) and

(i) (110) inverse fast Fourier transform (IFFT) image. g2 and i2 are the line maps along the red arrows in g and i, respectively.

The XRD results demonstrate the suppression of monoclinic (m-) phase in all three films (Figure S2). The W capping electrode with relatively low thermal expansion coefficient (TEC) compared to $ZrO_2$ and $HfO_2$ might be responsible for this structural evolution during the annealing process [22]. Electrodes that have relatively low TEC induce in-plane tensile strain during the cooling step of annealing process. The tensile strain may prevent the formation of m-phase by suppression of twin deformation [29]. The XRD patterns show the characteristic o- ($2\theta = 30.4°$) and t-phase ($2\theta = 30.7°$) peaks. To enable precise comparison, we obtained XRD patterns focused at a limited range in which the shape of o/t-phase characteristic peaks is most distinguishable (Figure 2b). The o/t XRD characteristic peak of HZZ nanolaminate lies between those of HZO and $ZrO_2$ thin films. The HZO and $ZrO_2$ thin films show symmetric peaks near $2\theta = 30.4°$ and $30.7°$ which indicate o-phase and t-phase dominant structures, respectively. The HZO device had a high $2P_r$, which represents the formation of a large fraction of o-phase. P-E measurement revealed a pure AFE behavior, which indicates a t-structure in $ZrO_2$ thin film.

The HZZ nanolaminate showed a broad and asymmetric peak at $2\theta = 30.5°$, which indicates the coexistence of o- and t-phases. These data confirmed a mixture of FE (from o-phase) and AFE (from t-phase) properties in the HZZ nanolaminate. The broad Bragg peak indicates that the nanolaminate pattern prevents the grain growth and results in a fine crystallite-structured thin film. The asymmetric pattern indicates coexisting t- and o-phase.

However, high resolution transmission electron microscopy (HRTEM) characterization would be a more practical approach to capture any local structural deviation in the HZZ phase caused by either o or t structures, or combination of them. For that, the HRTEM image of the HZZ nanolaminate along the [101] zone axis is shown in figure 2c, which advocates the well crystallized and homogenous structure of the annealed HZZ layer, surrounded by the two W electrodes. As it was discussed above, it is again worth mentioning that according to our previous study [29] the formation of a dead layer at W/HZZ interface is inevitable as the deposition process was carried out at 250 °C under 15-sec ozone pulses. Aygun *et al* [35] and Tan *et al* [36] studied the formation of an interfacial layer using the spectroscopic ellipsometer, XRD, Fourier transform infrared, and

XPS depth profiling techniques. In this work, the formation of interfacial layer could be observed directly through STEM and HRTEM analyses. Since the chemistry of $HfO_2$ and $ZrO_2$ materials are almost the same and the fabrication process of all devices were carried out at the same conditions, i.e. the ALD at 250 °C under a 15-sec ozone dosage and the RTA at 500 °C for 30 sec, therefore, we supposed that the formation of interfacial layers between W and $HfO_2$ or $ZrO_2$ are almost the same and the dielectric properties of each device were affected by this region equally.

The co-existence of the both o and t phases can be deducted based on the observation of extra (110) spots (red arrows) in the fast Fourier transform (FFT) pattern of the HZZ layer in figure 2d which can be further assessed as follows.

Firstly, in order to investigate the importance of the applied zone axis in HRTEM characterization, the selected area diffraction (SAED) simulation of $P$ca2$_1$ phase, along the [101] and [110] zone axes are shown in figure 2e and f, respectively [37]. Along the [110] axis in figure 2f both (001) and (1-10) spots are absent, which are indicated by yellow and red open circles, respectively. Along the [101] zone in figure 2e, (10-1) is forbidden. Therefore, in the case of either [110] or [101], the FFT pattern in figure 2d is clearly suggesting that we have another phase coexisting alongside with the commonly reported $P$ca2$_1$ phase, which is nothing but t-phase.

For more clarification, the (111) filtered inverse fast Fourier transform (IFFT) image represented in figure 2g. The line map captured along the red arrow well shows the absence of any kind of distortion in the (111) planes (Figure 2g$_2$). While a local distortion may be noticed along the white arrow in figure 2h when the same condition is governed for the (001) planes, these atomic planes have not experience any distortion, which implies the fact that the applied zone axis must be [101] as is shown in SAED simulation pattern in figure 2e. However, the situation is completely different for (110) planes in the HZZ layer. As is obvious in figure 2i, (110) atomic planes of the HZZ layer are extremely distorted along the drawn arrows, which is consistent with the captured line map along the red arrow in figure 2i$_2$.

To optimize the fabrication condition, the effect of post-annealing processes on the evolution of structural features, and the dielectric and FE properties was investigated.

*P-E* curves (Figure 3a) were obtained from HZZ nanolaminates annealed under different conditions. Annealing for 30 s at annealing temperature $T_{ANN}$ > 350 °C caused the appearance of

FE-AFE mixed loop. Below 350 °C the film showed a linear dielectric behavior (data was not presented here.).

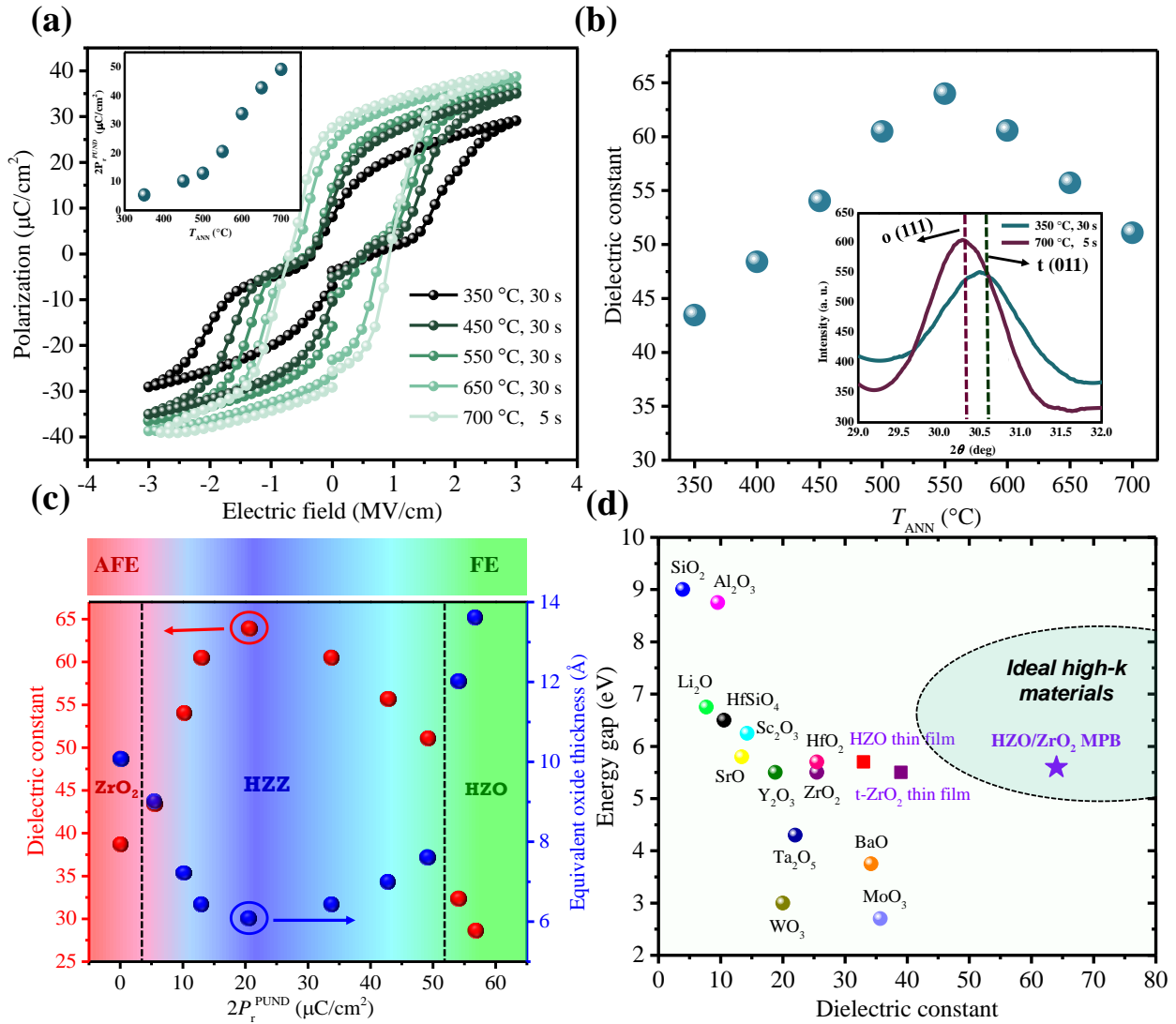

**Figure 3.** (a) P-E loops for the W/HZZ/W devices annealed at different temperatures $T_{ANN}$ for 30 s. Annealing was performed at 700 °C for 5 s to avoid aggressive reaction at the dielectric/metal interface. Inset: $2P_r^{PUND}$ for HZZ device annealed at different $T_{ANN}$. (b) Effect of $T_{ANN}$ on the dielectric constant of W/HZZ/W devices. (c) Dielectric constant (red spheres) and EOT (blue spheres) versus $2P_r^{PUND}$ values of $HfO_2$-$ZrO_2$ thin film with different design. (d) Experimentally-measured band gap versus dielectric constant of well-known oxides extracted from reference [5]. The present work (purple star) achieved an ideal

dielectric constant of HfO$_2$-ZrO$_2$ through MPB design. It was supposed that the nanolaminate design doesn't affect the band gap of HfO$_2$ and ZrO$_2$ materials.

Increase in $T_{ANN}$ widened the hysteresis loops, and caused an increase in 2$P_r$ and a gradual disappearance of the double-hysteresis characteristics. Therefore, AFE-dominant behavior changed to FE-dominant behavior as $T_{ANN}$ increased. 2$P_r$ substantially increased from 5.4 to 48.8 µC/cm$^2$ as $T_{ANN}$ was increased from 350 °C to 700 °C; the change can be attributed to an increase in the fraction of o-phase due to the intermixing of HfO$_2$ and ZrO$_2$ at the interfaces during the annealing process. Moreover, as $T_{ANN}$ is increased, the grain size increases, and may impede the stabilization of t-phase after cooling to RT. In fact, the o-phase is stable at medium grain sizes between t- and m-phases. The surface energy of o-phase is 2.575 J/cm$^2$, whereas it is 2.5 J/cm$^2$ for t-phase and 3.2 J/cm$^2$ for m-phase [38]. An increased in-plane tensile strain induced by the W top and bottom electrode can provide additional force to drive the transition from t-phase to o-phase by annealing at high $T_{ANN}$. In the XRD pattern of HZZ nanolaminates, the o/t characteristic peak of the structure was left-shifted after $T_{ANN}$ = 700 °C compared to $T_{ANN}$ = 350 °C (figure 3b); this change indicates an increase in the o-phase fraction throughout the film after annealing at $T_{ANN}$ = 700 °C. As $T_{ANN}$ increased, FWHM of the o/t characteristic peak decreased; the change is attributed to improved crystallization driven by the thermal energy. The t-phases are stable at the smallest crystallite sizes. Therefore, a combination of left-shift and broadened Bragg peak revealed that dominance of t-phase in the films increased as $T_{ANN}$ decreased.

A temperature at which the maximum $\rho_{MPB}$ can be achieved is promising for exceptionally high-$\kappa$ device. Increasing $T_{ANN}$ from 350 °C to 550 °C caused an increase in $\epsilon_r$. Annealing at 550 °C for 30 sec gave the maximum $\epsilon_r$ = 64, which may suggest that this condition yielded the maximum $\rho_{MPB}$ in HZZ nanolaminate. $T_{ANN}$ > 550 °C caused a reduction in $\epsilon_r$. Therefore, annealing at $T_{ANN}$ either higher than or lower than 550 °C caused a decrease in $\epsilon_r$ of the HZZ nanolaminate, possibly as a result of increase in formation of o- and t-phase, respectively. Consequently, the $\rho_{MPB}$ can decrease. 2$P_r$ were measured using positive-up-negative-down (PUND) method to correct for the effect of dielectric loss on the measured polarization. Increase in $T_{ANN}$ causes an increase in the leakage current of W/HZZ/W device by inducing formation of interfacial layers at the W/HZZ interface that can extract oxygen from HZZ layers [29]. The PUND method is helpful to subtract the effect of leakage current from the measured polarization. As 2$P_r$ increased, $\epsilon_r$ first increased to a maximum of 64 at $2P_r \approx 20$ µC/cm$^2$, then decreased (Figure 3c). This result may mean that $2P_r \approx 20$ µC/cm$^2$, induces the highest $\rho_{MPB}$ at the interfaces of HZO (FE)-ZrO$_2$ (AFE) nanolaminates. For comparison, the graph includes the dielectric properties of ZrO$_2$ ($2P_r = 0$) and HZO ($2P_r = 54$ µC/cm$^2$ (annealed at 500 °C for 30 s) and 57 µC/cm$^2$ (annealed at 700 °C for 5 s)) at the ends. The ZrO$_2$ thin film with $2P_r = 0$ gives a $\epsilon_r$ of 39 and a pure HZO film shows $28 \leq \epsilon_r \leq 32$ depending on the annealing conditions. A nanolaminate structure annealed at an appropriate condition had a dielectric constant 73% higher than that of ZrO$_2$ and HZO films. The EOT measurements showed a reduction to ~ 6 Å. Assuming that the nanolaminate design does not affect $E_g$ of either parent phases (ZrO$_2$, HfO$_2$) then $E_g$ as a function of $\epsilon_r$ (figure 3d) shows that the new design is a highly promising approach towards the production of ideal high-$\kappa$ materials.

**Conclusion**

The dielectric properties of HfO$_2$-ZrO$_2$ based thin film was enhanced through HZZ nanolaminate design. While the HZO and ZrO$_2$ thin films showed pure FE and AFE features, respectively, the HZZ nanolaminate revealed a complex P-E loop consists of both FE and AFE characteristics. The unusual large $\epsilon_r$ of HZZ nanolaminate compared to the HZO and ZrO$_2$ films was attributed to the formation of MPB region. In fact, the P-E curve showed a combination of FE and AFE characteristics which may introduce MPB regions throughout the nanolaminate structure. The coexistence of t and o phases was confirmed through XRD and HRTEM studies and the nature of nanolaminate structure was confirmed by STEM analysis. To increase the formation of MPB region inside the HZZ nanolaminate, the annealing condition was adjusted. A high $\epsilon_r$ of 64 was achieved after an appropriate annealing process. This work presents a new design to increase the dielectric properties of HfO$_2$-ZrO$_2$ based thin film. The fabrication process is performed at a low thermal budget (250 °C) and the device is compatible with silicon technology, so the new design introduces a promising device for future electronics.

**Acknowledgements**

The authors thank Mr. Mehrdad Ghiasabadi Farahani, Dr. Stanislav Kamba, Dr. Kai Ni, Dr. Writam Banerjee, Dr. Nikam and Mr. Peyman Asghari-Rad for helpful discussions. This work was supported by the National Research Foundation of Korea funded by the Korean government (MSIT), grant no. NRF-2018R1A3B1052693.

## Contributions

A. Kashir conceived the idea, performed the experiments, analyzed the results and wrote the manuscript. H. Hwang supervised the project and reviewed the manuscript.